\documentclass[aps,nofootinbib,preprint]{revtex4}
\usepackage{amsfonts,amssymb,amsmath}
\usepackage{graphicx}
\usepackage{subfigure}
\usepackage{wrapft}

\newcommand{\f}[2]{\frac{#1}{#2}}
\newcommand{\mk}[1]{\left( #1 \right)}

\newcommand{\be}{\begin{equation}}
\newcommand{\ee}{\end{equation}}
\newcommand{\lcdm}{{\rm \Lambda CDM}}
\newcommand{\frg}{{\rm fRG}}
\newcommand{\eff}{{\rm eff}}

\begin{document}

\title{
  Phantom behaviour and growth index anomalous evolution
  in viable $f(R)$ gravity models
}
\author{
  Hayato Motohashi\footnote{Email address: motohashi@resceu.s.u-tokyo.ac.jp}$^{(a),(b)}$
  ,
  Alexei A. Starobinsky\footnote{Email address: alstar@landau.ac.ru}$^{(b),(c)}$
  and
  Jun'ichi Yokoyama\footnote{Email address: yokoyama@resceu.s.u-tokyo.ac.jp}$^{(b),(d)}$
}
\address{
$^{(a)}$ Department of Physics, Graduate School of Science,
The University of Tokyo, Tokyo 113-0033, Japan \\
$^{(b)}$ Research Center for the Early Universe (RESCEU), 
Graduate School of Science, The University of Tokyo, Tokyo 113-0033, Japan \\
$^{(c)}$ L. D. Landau Institute for Theoretical Physics, 
Moscow 119334, Russia \\
$^{(d)}$ Institute for the Physics and Mathematics of the Universe(IPMU), 
The University of Tokyo, Kashiwa, Chiba, 277-8568, Japan
}

\begin{abstract}
We present numerical calculation of the evolution of a background
space-time metric and sub-horizon matter density perturbations in
viable $f(R)$ gravity models of present dark energy and cosmic
acceleration. We found that viable models generically exhibit
recent crossing of the phantom boundary $w_{\rm DE}=-1$. Moreover, as
a consequence of the anomalous growth of density perturbations
during the end of the matter-dominated stage, their growth index
evolves non-monotonically with time and may even become negative.
\end{abstract}

\begin{flushright}
RESCEU-26/10
\end{flushright}

\maketitle

\section{Introduction}
It is one of the most important issues for cosmologists and
particle physicists to understand the physical origin of the dark
energy (DE) which is responsible for an accelerated expansion of
the current Universe. Although the standard spatially flat ${\rm
\Lambda}$-Cold-Dark-Matter ($\lcdm$) model is consistent with all
kinds of current observational data \cite{O01_WMAP7},
some tentative deviations from it have been reported recently
\cite{O01_Shafieloo:2009ti,O01_Bean:2009wj}. Furthermore, in the
$\lcdm$ model, the cosmological term is regarded as a new
fundamental constant whose observed value is much smaller than any
other energy scale known in physics. Hence it is natural to seek
for non-stationary models of the current DE. Among them, $f(R)$
gravity which modifies and generalizes the Einstein gravity by
incorporating a new phenomenological function of the Ricci scalar
$R$, $f(R)$, can provide a self-consistent and non-trivial
alternative to the $\lcdm$
model\cite{O01_Hu:2007nk,O01_Starobinsky:2007hu}.

In the previous paper \cite{O01_Motohashi:2009qn}, we calculated
evolution of matter density fluctuations in viable $f(R)$ models
\cite{O01_Hu:2007nk,O01_Starobinsky:2007hu} for redshifts $z \gg
1$ during the matter-dominated stage and found an analytic
expression for them. In this paper we extend the previous analysis
and perform numerical calculations of the evolution of both
background space-time and density fluctuations for the particular
$f(R)$ model of Ref.~\cite{O01_Starobinsky:2007hu} without such a
restriction. As a result, we have found crossing of the phantom
boundary $w_{\rm DE}=-1$ at an intermediate redshift $z\lesssim 1$ for
the background space-time metric and an anomalous behavior of the
growth index of fluctuations.

\section{Background}
We adopt the following action 
of $f(R)$ models with model parameters $n,~\lambda$ and $R_s$
\cite{O01_Starobinsky:2007hu}: 
\be \label{O01_fR} S=
\frac{1}{16\pi G} \int d^4x \sqrt{-g} f(R) + S_m, \quad f(R)=R +
\lambda R_s \left[ \left( 1 +
\frac{R^2}{R_s^2}\right)^{-n}-1\right] , \ee 
where $S_m$ is the
action of the matter content which is assumed to be minimally
coupled to gravity. To make the late-time asymptotic de Sitter
regime where $R=$ constant stable, $\lambda$ has to satisfy
$f'(R)>Rf''(R)$. As a result, $\lambda$ has a lower limit
$\lambda_{\min}$ for each $n$. Numerically we find
$(n,\lambda_{\min})=$(2, 0.9440), (3, 0.7259), and (4, 0.6081).
From the action \eqref{O01_fR}, we obtain field equations as
\begin{align}
 R^{\mu}_{\nu}-\frac{1}{2}\delta^{\mu}_{\nu}R&=
-8\pi G\mk{T^{\mu}_{\nu (m)}+T^\mu_{\nu ({\rm DE})}}, \\
\label{O01_EMtensor}
8\pi G T^\mu_{\nu ({\rm DE})}&\equiv
(F-1)R^\mu_\nu-\frac{1}{2}(f-R)\delta^{\mu}_{\nu}
+(\nabla^\mu\nabla_\nu-\delta^{\mu}_{\nu}\square)F,\quad
F(R)\equiv f'(R).
\end{align}
Working in the spatially flat Friedmann-Robertson-Walker (FRW)
space-time with a scale factor $a(t)$,
\begin{align}
\label{O01_hubble} 3H^2&=8\pi G\rho-3(F-1)H^2+\frac{1}{2}(FR-f)-3H\dot F, \\
\label{O01_hdot} 2\dot{H}&=-8\pi G\rho -2(F-1)\dot{H}-\ddot{F}+H\dot{F},
\end{align}
where $H$ is the Hubble parameter and $\rho$ is the energy density
of the material content which we assume to consist of
non-relativistic matter. From \eqref{O01_fR}, we can determine the
DE equation of state parameter $w_{\rm DE}$, 
\be \label{O01_wDE}
w_{\rm DE}\equiv\f{P_{\rm DE}}{\rho_{\rm DE}}
=-1+\f{2\dot{H}(F-1)-H\dot{F}+\ddot{F}}{-3H\dot{R}F'
+3(H^2+\dot{H})(F-1)-(f-R)/2}. \ee

We solve the evolution equation \eqref{O01_hdot} numerically using
\eqref{O01_hubble} to check numerical accuracy. The moment when
the matter density parameter $\Omega(t)=16\pi G\rho/(16\pi
G\rho+\lambda R_s)$ equals to $0.998$ is chosen as the initial
time $t_i$. We determine the current epoch $t=t_0$ by the
requirement that the value of $\Omega$ takes the observed central
value $\Omega_0=0.27$. $R_s$ is fixed in such a way as to
reproduce the current Hubble parameter $H_0=72$km/s/Mpc. We find
the ratio $R_s/H_0^2$ is well fit by a simple power-law
$R_s/H_0^2=c_n\lambda^{-p_n}$ with $(n,c_n,p_n)=$(2, 4.16, 0.953),
(3, 4.12, 0.837), and (4, 4.74, 0.702), respectively, whereas in
the $\lcdm$ limit it would behave as
$R_s/H_0^2=6(1-\Omega_0)/\lambda\simeq 4.38\lambda^{-1}$.

\begin{figure}[t]
\centering
{
\includegraphics[width=80mm]{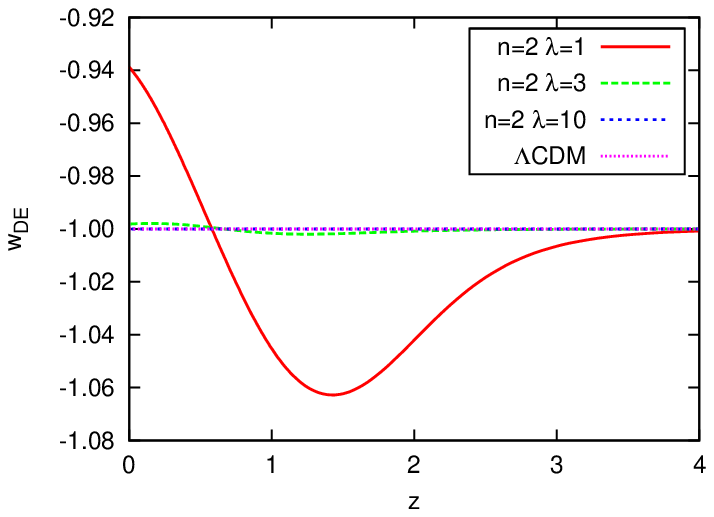}
\label{O01_fig:w-a}}
{
\includegraphics[width=80mm]{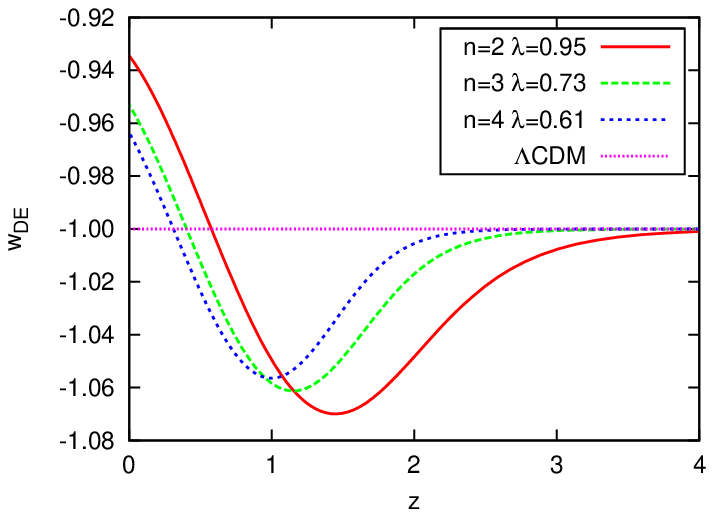}
\label{O01_fig:w-b}}
\caption{Evolution of the equation-of-state parameter of
effective dark energy.}
\label{O01_fig:w}
\end{figure}

Figures \ref{O01_fig:w} depict evolution of $w_{\rm DE}$ as a
function of redshift $z$ where phantom crossing is manifest. As
expected, it approaches $w_{\rm DE}=-1=\text{constant}$ as we
increase $\lambda$ for fixed $n$. For the minimal allowed values
of $\lambda$, deviations from $w_{\rm DE}=-1$ are observed at 5\%
level on both directions in $z \lesssim 2$ independently of $n$.
From \eqref{O01_wDE}, we can read off that this phantom crossing
behavior is not peculiar to the specific choice of the function
\eqref{O01_fR} but a generic one for models which satisfy the
stability condition $F'>0$.

\section{Perturbations}
We now turn to the evolution of density fluctuations. In $f(R)$
gravity, the evolution equation of density fluctuations, $\delta$,
deeply in the sub-horizon regime is given
by\cite{O01_Tsujikawa:2007gd} 
\be \label{O01_de1} \ddot \delta +
2H\dot \delta - 4\pi G_{\text{eff}}\rho \delta = 0, \quad
G_{\text{eff}}=\frac{G}{F} \frac{1+4\frac{k^2}{a^2}\frac{F'}{F}}
{1+3\frac{k^2}{a^2}\frac{F'}{F}}. \ee 
In the previous
paper\cite{O01_Motohashi:2009qn} we obtained an analytic solution
for the high-curvature regime when the scale factor evolves as
$a(t)\propto t^{2/3}$ and $F$ takes an asymptotic form $F\simeq
1-2n\lambda \mk{R/R_s}^{-2n-1}$. In the present paper, we
numerically integrate \eqref{O01_de1} up to $z=0$ without using
the approximation $|F-1|\ll 1$.

The wavenumber of our particular interest is the scale
corresponding to $\sigma_8$ normalization, for which we find
$k_\eff(r=8h^{-1}{\rm Mpc})=0.174h{\rm Mpc}^{-1}$. Since the
standard $\lcdm$ model normalized by CMB data explains galaxy
clustering at small scales well, $\delta_{\frg}$ should not be too
much larger than $\delta_\lcdm$ on these scales. We may typically
require $(\delta_{\frg}/\delta_{\lcdm})^2\lesssim 1.1$. Although
we neglect non-linear effects here, the difference between linear
calculation and non-linear N-body simulation remained smaller than
5\% at wavenumber $0.174h{\rm Mpc}^{-1}$\cite{O01_Oyaizu:2008tb}.

\begin{figure}[tbp]
\centering
{
\includegraphics[width=80mm]{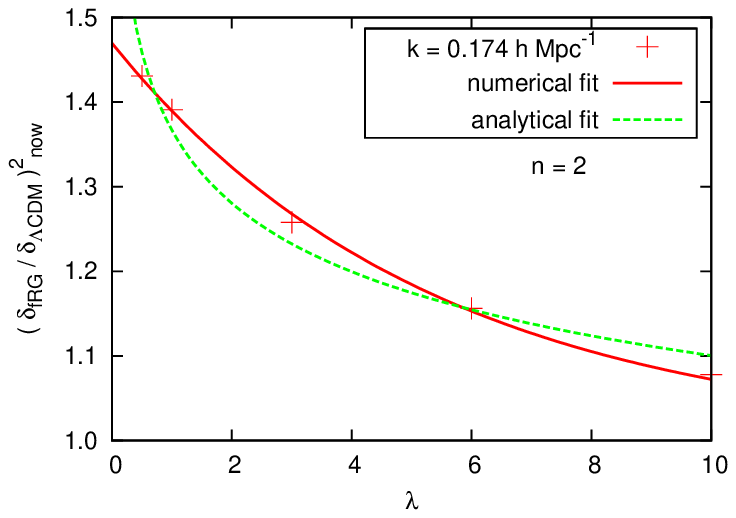}
\label{O01_fig:rtnl-a}}
{
\includegraphics[width=80mm]{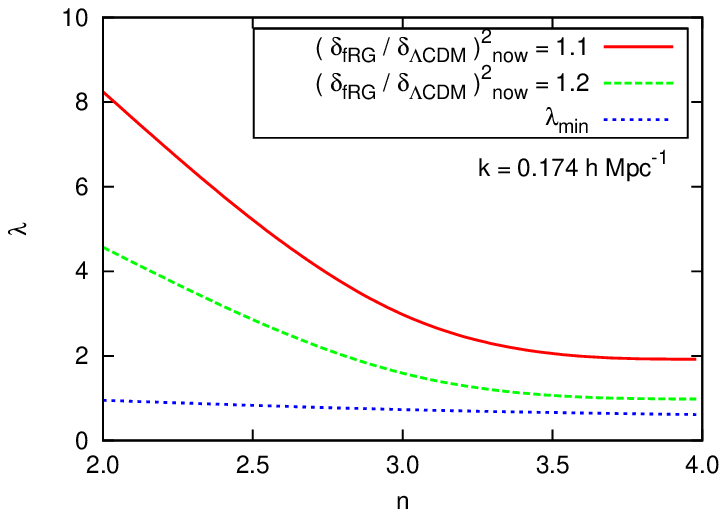}
\label{O01_fig:rtnl-b}} \caption{Constraints from the power
spectrum.} \label{O01_fig:rtnl}
\end{figure}

The left panel of Fig.~\ref{O01_fig:rtnl} represents
$(\delta_{\frg}/\delta_{\lcdm})^2$ as a function of $\lambda$ for
$n=2$ together with two fitting functions. The solid line is from
the analytic formula obtained in Ref.~\cite{O01_Motohashi:2009qn},
and the broken line is numerical fitting using an exponential
function $1+b_ne^{-q_n\lambda}$. From these analysis, we can
constrain the parameter space as the right panel of Fig.~\ref{O01_fig:rtnl}. 
The region which satisfy $(\delta_{\frg}/\delta_{\lcdm})^2 < 1.1$ lies
above the solid line. The region below the dotted line is
forbidden from instability of de Sitter regime.

Next we turn to another important quantity used to distinguish
different theories of gravity, namely, the gravitational growth
index, $\gamma(z)$, of density fluctuations. It is defined through
\be \f{d\ln\delta}{d\ln a}=\Omega_m(z)^{\gamma(z)},~~~\text{or}~~~
  \gamma(z)=\f{\log\mk{\f{\dot{\delta}}{H\delta}}}{\log\Omega_m}. \ee
It takes a practically constant value $\gamma\cong 0.55$ in the
standard $\lcdm$ model,
but it evolves with time in modified
gravity theories in general.  We also note that $\gamma(z)$ has a
nontrivial $k$-dependence in $f(R)$ gravity since density
fluctuations with different wavenumbers evolve differently.
Therefore, this quantity is a useful measure to distinguish
modified gravity from $\lcdm$ model in the Einstein gravity.

\begin{figure}[t]
\centering
{
\includegraphics[width=80mm]{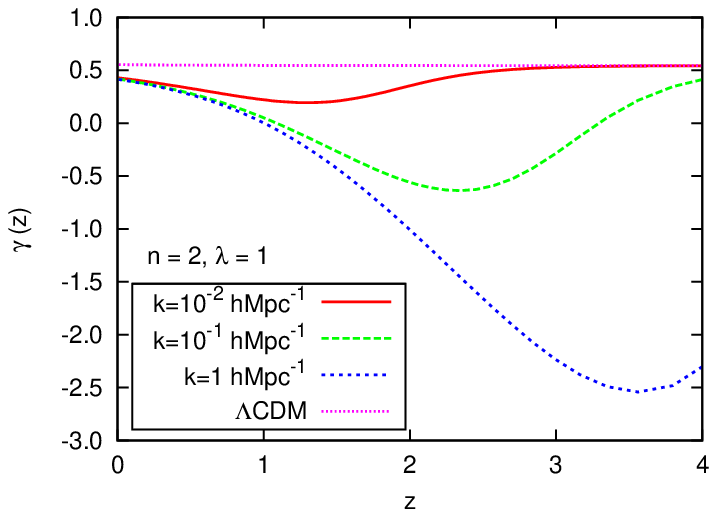}
\label{O01_fig:gG-a}}
{
\includegraphics[width=80mm]{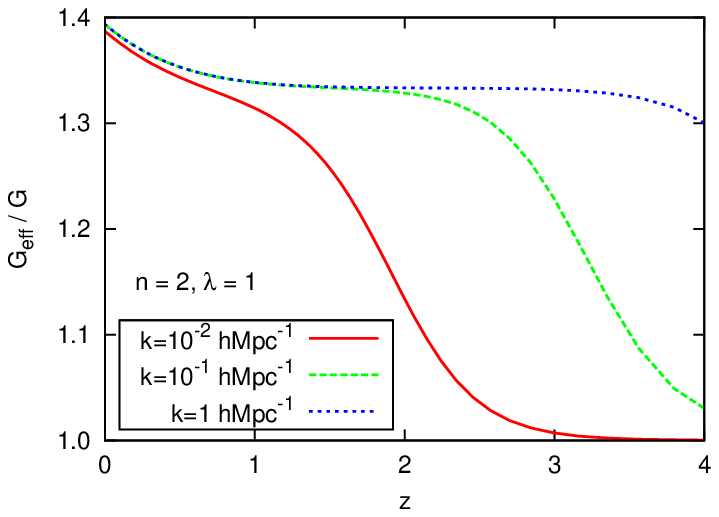}
\label{O01_fig:gG-b}}
\caption{Evolution of $\gamma(z)$ and $G_\eff/G$.}
\label{O01_fig:gG}
\end{figure}

Figures \ref{O01_fig:gG} show the evolution of $\gamma(z)$ together
with that of $G_\eff/G$ for different values of $k$. In the early
high-redshift regime, $\gamma(z)$ takes a constant value identical
to the $\lcdm$ model. It gradually decreases in time, reaches a
minimum which may be even negative, and then increase again
towards the present epoch. We note that recently Narikawa and
Yamamoto\cite{O01_Narikawa:2009ux} numerically calculated time
evolution of $\gamma(z)$ in a simplified model which we had used
in the previous paper and also obtained some analytic expansion,
which behaves qualitatively the same as the numerical result but
with much more exaggerated amplitudes.  Our results, which satisfy
all viability conditions, exhibit milder deviation from $\lcdm$
model than they found. At present, the constraints for the growth
index is not so strict to distinguish the deviation from the
$\lcdm$ model\cite{O01_Rapetti:2009ri}, but observations may
reveal its time and wave number dependence in future.

\section{Conclusion}
In the present paper we have numerically calculated the evolution
of both homogeneous background and density fluctuations in a
viable $f(R)$ model of accelerated expansion based on the specific
functional form proposed in Ref.\ \cite{O01_Starobinsky:2007hu}.
We have found that viable $f(R)$ gravity models of accelerated
expansion generically exhibit phantom behavior during the
matter-dominated stage with crossing of the phantom boundary
$w_{\rm DE}=-1$ at redshifts $z\lesssim 1$. The predicted time
evolution of $w_{\rm DE}$ has qualitatively the same behaviour as
that was recently obtained from observational data in
\cite{O01_Shafieloo:2009ti}.

As for density fluctuations, we have numerically confirmed our
previous analytic results on a shift in the power spectrum index
for large wavenumbers which exceed the scalaron mass during the
matter dominated epoch\cite{O01_Motohashi:2009qn}, while for
smaller wavenumbers fluctuations have the same amplitude as in the
$\lcdm$ model.

We have also investigated the growth index $\gamma(k,z)$ of
density fluctuations and have given an explanation of its
anomalous evolution in terms of time dependence of $G_\eff$. Since
$\gamma$ has characteristic time and wavenumber dependence, future
detailed observations may yield useful information on the validity
of $f(R)$ gravity through this quantity, although current
constraints have been obtained assuming that it is constant both
in time and in
wavenumber\cite{O01_Bean:2009wj,O01_Rapetti:2009ri}.

\acknowledgments{
HM and JY are grateful to T.\ Narikawa and K.\ Yamamoto for useful
communications.
AS acknowledges RESCEU hospitality as a visiting professor. He was also
partially supported by the grant RFBR 08-02-00923 and by the Scientific
Programme ``Astronomy'' of the Russian Academy of Sciences.
This work was supported in part by
JSPS Grant-in-Aid for Scientific Research No.\ 19340054(JY),
JSPS Core-to-Core program  ``International Research Network
on Dark Energy'', and
Global COE Program ``the Physical Sciences Frontier'', MEXT, Japan.
}


\begin{thebibliography}{99}
\bibitem{O01_WMAP7} 
  E.~Komatsu {\it et al.},
  \href{http://jp.arxiv.org/abs/1001.4538}{arXiv:1001.4538}.
\bibitem{O01_Shafieloo:2009ti}
  A.~Shafieloo, V.~Sahni and A.~A.~Starobinsky,
  Phys.\ Rev.\  D {\bf 80}, 101301 (R) (2009)
  [arXiv:0903.5141].
\bibitem{O01_Bean:2009wj}
  R.~Bean,
  arXiv:0909.3853.
\bibitem{O01_Hu:2007nk}
  W.~Hu and I.~Sawicki,
  Phys.\ Rev.\  D {\bf 76}, 064004 (2007)
  [arXiv:0705.1158].
\bibitem{O01_Starobinsky:2007hu}
  A.~A.~Starobinsky,
  JETP Lett.\  {\bf 86}, 157 (2007)
  [arXiv:0706.2041].
\bibitem{O01_Motohashi:2009qn}
  H.~Motohashi, A.~A.~Starobinsky and J.~Yokoyama,
  Int.\ J.\ Mod.\ Phys.\ D {\textbf 18}, 1731 (2009)
  [arXiv:0905.0730].
\bibitem{O01_Tsujikawa:2007gd}
  S.~Tsujikawa,
  Phys.\ Rev.\  D  {\bf 76}, 023514 (2007)
  [arXiv:0705.1032].
\bibitem{O01_Oyaizu:2008tb}
  H.~Oyaizu, M.~Lima and W.~Hu,
  Phys.\ Rev.\  D {\bf 78}, 123524 (2008)
  [arXiv:0807.2462].
\bibitem{O01_Narikawa:2009ux}
  T.~Narikawa and K.~Yamamoto,
  arXiv:0912.1445.
\bibitem{O01_Rapetti:2009ri}
  D.~Rapetti, S.~W.~Allen, A.~Mantz and H.~Ebeling,
  arXiv:0911.1787.
\end{thebibliography}
\end{document}